\documentclass[twoside]{article}
\usepackage{fleqn,espcrc2}

\usepackage{graphicx}
\usepackage[figuresright]{rotating}

\newcommand{\AmS}{{\protect\the\textfont2
  A\kern-.1667em\lower.5ex\hbox{M}\kern-.125emS}}

\title{Nonsingular vortices in (s+d)-wave superconductors}
\author{A. S. Mel'nikov, I. M. Nefedov, D. A. Ryzhov,
I. A. Shereshevskii, P. P. Vysheslavtsev
        \thanks{This work was supported by the Russian Foundation for
        Basic Research (Grant No. 99-02-16188).}
\address{Institute for Physics of Microstructures,
        Russian Academy of Sciences\\
        603600, Nizhny Novgorod, GSP-105, Russia}}

\begin{document}
\pagestyle{empty}
\begin{abstract}
The structure of a single flux line in (s+d)-wave superconductors has
been analyzed within the Ginzburg-Landau (GL) model generalized for two
order parameter components.
The fourfold symmetric singular vortex solution is shown to be unstable
in a certain range of the GL parameters with respect to the mutual shift
of s- and d- wave unit vortices.
The resulting nonsingular vortex structure is studied both analytically and
numerically.
\vspace{1pc}
\end{abstract}

\maketitle

Recently the distinctive characteristics of vortices in unconventional
superconductors which can be described by the phenomenological
Ginzburg-Landau (GL) theory with a multicomponent order parameter (OP)
are of great interest in connection with the investigations of the mixed
state structure in high-$T_c$ and heavy fermion compounds which are
strong candidates for unconventional superconductivity.
A single flux line in such systems is known to contain a set of unit
vortices of different OP components
\cite{Sauls,Barash,Berlinsky,Sigrist}.
One can indentify two possible types of flux lines: (i)~singular vortices
(which have at least one point where the superconducting gap is zero) and
(ii)~nonsingular vortices (where the gap is nonzero everywhere in the
vortex core).
In this paper we focus on the case of ${s+d_{x^2-y^2}}$-wave pairing
(which can be relevant to the case of high-$T_c$ superconductors) and
consider the range of GL functional parameters where the singular
vortices (studied in \cite{Barash,Berlinsky,Sigrist}) become unstable
according to the scenario analogous to the one proposed in
\cite{Sauls,Barash} for heavy fermion compounds.
The goal of this paper is to analyse the detailed structure of resulting
nonsingular vortices using both numerical and analytical methods.

\begin{figure}[htb]
\leavevmode
\includegraphics[width=50mm]{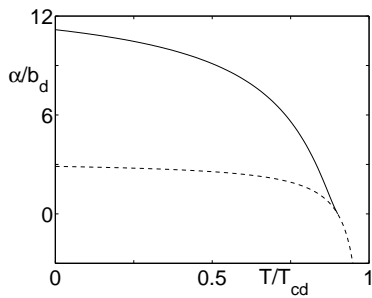}
\caption{The phase diagram on $\alpha-T$-plane for
        ${\alpha_s=3.2\alpha_d}$, ${T_{cs}=0.9T_{cd}}$,
	${b_s=9b_d}$, ${\beta=0}$, ${K_s=0.5K_d}$.
        The homogeneous states with ${\Psi_s=0}$, ${\Psi_d\neq 0}$ and
        with ${\Psi_s,\Psi_d\neq 0}$ exist above and below the dashed
        line, respectively.
        The nonsingular vortices exist between the dashed line and the
        solid line.}
\label{fig:diagram}
\end{figure}

\begin{figure}[htb]
\leavevmode
\includegraphics[width=75mm]{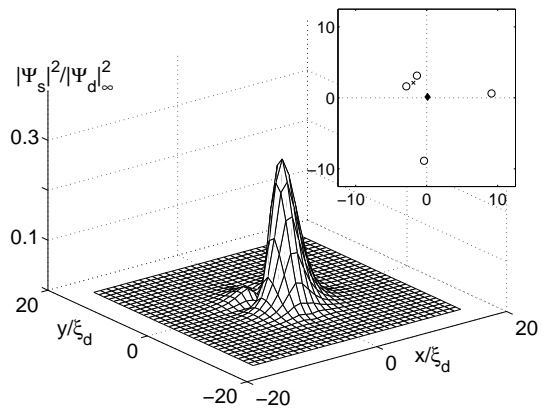}
%
\caption{The typical structure of s-wave OP for nonsingular vortex
        for ${\alpha_s=3.2\alpha_d}$, $T_{cs}=0.95T_{cd}$, $T=0.68T_{cd}$,
	$b_s=9b_d$, $\alpha=3b_d$, $\beta=0$, $K_s=0.5K_d$,
	$\gamma=0.65K_d$.
        The inset shows the positions of d-wave vortex whose winding
        number ${N=+1}$ (black diamond), the s-wave vortices with
        ${N=+1}$ (circles) and with ${N=-1}$ (cross).}
\label{fig:structure}
\end{figure}

We start with the GL free energy functional generalized for two
components of the OP $\Psi_d$ and $\Psi_s$ corresponding to the
$d_{x^2-y^2}$-wave and s-wave pairing, respectively \cite{Sigrist}:
$$
F=\int\{a_d|\Psi_d|^2+a_s|\Psi_s|^2
+\frac{b_d}{2}|\Psi_d|^4
+\frac{b_s}{2}|\Psi_s|^4
\qquad
$$
$$
+\alpha|\Psi_d|^2|\Psi_s|^2
+\frac{\beta}{2}(\Psi_d^{2}\Psi_s^{*2}
+\Psi_{d}^{*2}\Psi_{s}^{2})
\qquad\qquad\qquad
$$
$$
+K_s|{\bf\Pi}\Psi_s|^2
+K_d|{\bf\Pi}\Psi_d|^2
+\gamma[(\Pi_x^*\Psi_s^*\Pi_x\Psi_d
\qquad\qquad\qquad
$$
\begin{equation}
\label{main1}
-\Pi_y^*\Psi_s^*\Pi_y\Psi_d)+c.c.]
+\frac{{\bf H}^2}{8\pi}\}\,d{\bf r},
\end{equation}

\noindent
where ${{\bf \Pi}=\nabla-i\frac{2\pi}{\Phi_0}{\bf A}}$,
${{\bf H}=curl{\bf A}}$, ${{\bf r}=(x,y)}$, ${a_s=\alpha_s (T-T_{cs})}$,
${a_d=\alpha_d (T-T_{cd})}$ (we assume ${T_{cs}<T_{cd}}$ and the magnetic
field is applied parallel to the $c$-axis).
If the GL parameters are chosen so that there is an one-component
homogeneous state (${{\bf H}=0}$), then for ${{\bf H} \neq 0}$ the
subdominant OP can appear in the vortex core regions with the
inhomogeneous dominant OP component either due to the mixing gradient
terms (${\gamma \neq 0}$) \cite{Berlinsky,Sigrist} or due to the
instability of the inhomogeneous state against the formation of the
subdominant OP nucleus \cite{Sauls,Barash,Laughlin}.
Let us consider the latter mechanism (which is obviously responsible for
the formation of nonsingular vortices) for the simplest case
${\gamma=0}$, $\beta=0$ and continue with the analysis of the linearised
GL equation of $\Psi_s$:

\begin{equation}
\label{main2}
-K_s\nabla^2\Psi_s+\alpha|\Psi_d|^2\Psi_s=-a_s\Psi_s,
\end{equation}

\noindent
where $\Psi_d$ describes the vortex solution within the conventional
single-component GL theory, also we neglect~${\bf A}$.
The "lowest energy eigenstate" of the Schr\"odinger equation
(\ref{main2}) defines the temperature~$T^*$ of the phase transition into
nonsingular vortex state.
Using the following approximation
${|\Psi_d|=|\Psi_d|_{\infty}r/\sqrt{r^2+2{\xi_d}^2}}$ (here
$|\Psi_d|_{\infty}=\sqrt{|a_d|/b_d}$, ${\xi_d=\sqrt{K_d/|a_d|}}$) for the
cases ${K_s\ll K_da_s/a_d}$ and ${\alpha\ll b_dK_s/K_d}$, we obtain
${-a(T^*)=\sqrt{2K_s\alpha/(K_db_d)}}$ and ${-a(T^*)\simeq \alpha/b_d}$,
respectively (here ${a(T)=a_s(T)/|a_d(T)|}$).
We propose the following extrapolation for the phase transition curve on
$\alpha-T$ plane (Fig.\ref{fig:diagram}), which separates the regions
with singular and nonsingular vortices:
${\alpha=-b_da(T^*)(1-a(T^*)K_d/(2K_s))}$.
This expression is valid for the above two limiting cases and for the
particular case (${\alpha_s=1.25\alpha_d}$, ${T=0}$, $T_{cs}=0.8T_{cd}$,
${b_s=b_d}$, ${\beta=0}$, ${K_s=K_d}$, ${\gamma=0}$) studied in \cite{Sauls}.
The analysis of the structure of nonsingular vortices has been carried out
using numerical calculations based on the time-dependent GL theory.
It was found out that for ${\gamma\neq 0}$ the nonsingular vortices with
broken fourfold symmetry (see Fig.\ref{fig:structure}) are stable below
the phase transition curve (solid line in Fig.\ref{fig:diagram}), whereas
above the curve it is the singular fourfold symmetric vortices that are
energetically favourable.

In conclusion, we have considered the structure of nonsingular vortices
in ${s+d_{x^2-y^2}}$ superconductor and found the range of parameters
where these vortices exist.
The nonsingular vortices in such materials have neither the fourfold
symmetry, nor the normal core region.
This can lead to some interesting effects (e.g. nontrivial dynamics,
moderate pinning effects etc.).

\end{document}